\documentclass{article}
\usepackage{epsfig}
\def\hei{\hbox{He\,{\sc i}}} 
\def\heii{\hbox{He\,{\sc ii}}}
\newcommand{\vsini}{\mbox{$v\sin\!i$}}
\newcommand{\logg}{\mbox{$\log g$}}
\newcommand{\teff}{\mbox{$T_{\mathrm{eff}}$}} 
\begin{document} 
\LARGE{\bf {The helium abundances in HgMn and normal 
stars}}\footnote{Poster paper presented at IAU Symposium 224, ``The 
A-Star Puzzle'', Poprad, Slovakia, 2004 July 7-13}
\\

\Large { 

\noindent M. M. Dworetsky\\
{\em Department of Physics \& Astronomy, University College 
London, 
Gower St., London WC1E 6BT, UK mmd@star.ucl.ac.uk}}\\

\large

{\bf {Abstract.}} The parameter-free model of diffusion in the atmospheres
of HgMn stars (Michaud 1986; Michaud et al 1979)  predicts that helium
should sink below the \heii\ ionization zone in order that diffusion of
other elements may take place, and that all HgMn stars
should have deficits of helium in their photospheres, with a minimum
deficit of 0.3 dex.  In this study, the Smith \& Dworetsky (1993) sample
of HgMn stars and normal comparison stars is examined, and the helium
abundances determined by spectrum synthesis using \'{e}chelle spectra
taken at Lick Observatory and the AAT.  The prediction is confirmed; all
HgMn stars are deficient in He by as much as 1.5 dex.  Also, two HgMn
stars, HR7361 and HR7664, show clear evidence of helium stratification.

\vspace{0.5cm}

{\bf {Introduction.}} Abundances were determined for 25 HgMn stars and 12
normal and su\-per\-fic\-ial\-ly normal stars of similar \teff, using an
LTE analysis.  It is well-known that the effects of non-LTE can safely be
ignored in the relevant temperature range.  The analysis was performed for
two \hei\ lines, $\lambda$4026.2\ and $\lambda$4471.5. The line profile
tables of Barnard et al (1969, 1974, 1975) and Shamey (1969) were used.  
The abundances and estimated errors were obtained here by trial and error
fits by eye to the observations.  The unweighted mean for the normal stars
is $\log N(\mathrm {He})/N(\mathrm H) = 10.98 \pm 0.05$, in excellent
agreement with the standard value 10.99 (Grevesse et al 1996).  (In this
paper all abundances are given on the scale $\log N(\mathrm H) = 12.00.$)  
It is found that all HgMn stars have underabundances, ranging from factors
of 0.3 dex at low \teff\ to 1.5 dex at high \teff.  These observations
provide direct support for the parameter-free model.

\vspace{0.3cm}

\begin{table}
\caption{Stellar Parameters and Helium Abundances for the 
Programme Stars.}
\begin{tabular}{lrrrccrrr}
\hline
Star &HD &\teff &\logg          &$\xi$       &\vsini\ & $\log N(\mathrm 
{He})$ & $\log N(\mathrm {He})$ & $\log N(\mathrm {He})$ \\
     &   & (K)   &(cm s$^{-2}$)& km s$^{-1}$ & km 
s$^{-1}$ & $\lambda$4471\phantom{00} & $\lambda$4026\phantom{00} 
& mean\phantom{001} \\
\hline
\multicolumn{9}{c}{Normal Stars}\\
\hline
$\nu$ Cap & 193432 & 10300 & 3.90 & 1.6 & 27 & $10.97\pm0.05$ & 
$10.70\pm0.10$ & $10.92\pm0.05$ \\
$\alpha$ Lyr & 172167 &9450 
&4.00&2.0&24&$10.99\pm0.05$&$10.99\pm0.10$&$10.99\pm0.05$ \\
HR7098&174567&10200&3.55&1.0 
&11&$10.90\pm0.05$&$10.50\pm0.05$&$10.70\pm0.04$ \\
$\zeta$ Dra & 155763 & 12900 & 3.90 & 2.5: & 34 & $10.99\pm0.05$ & 
$10.99\pm0.05$ & $10.99\pm0.04$ \\
134 Tau & 38899 & 10850 & 4.10 & 1.6 & 30 & $11.05\pm0.05$ & 
$10.99\pm0.10$ & $11.04\pm0.05$ \\
$\xi$ Oct & 215573 & 14050 & 3.85 & 0.5: & 5 & $10.99\pm0.10$ 
&$10.99\pm0.10$ &$10.99\pm0.07$ \\
$\tau$ Her & 147394 & 15000 & 3.95 & 0.0 & 32 & $10.95\pm0.05$ & 
$10.92\pm0.10$ & $10.94\pm0.05$ \\
21 Aql & 179761 & 13000 & 3.50 & 0.2 & 17 & $11.10\pm0.05$ & 
$11.00\pm0.10$ & $11.08\pm0.05$ \\
$\pi$ Cet & 17081 & 13250 & 3.80 & 0.0 & 25 & $11.20\pm0.05$ & 
$10.99\pm0.10$ & $11.16\pm0.05$ \\
\hline
\multicolumn{9}{c}{Superficially Normal Stars}\\
\hline
21 Peg & 209459 & 10450 & 3.50 & 0.5 & 4 & $10.90\pm0.05$ & 
$10.85\pm0.05$ & $10.88\pm0.04$ \\
HR7878 & 196426 & 13050 & 3.85 & 1.0: & 6 & $10.99\pm0.05$ & 
$10.85\pm0.10$ & $10.96\pm0.05$ \\
HR7338 & 181470 & 10250 & 3.75 & 0.5 & 3 & $10.80\pm0.10$ & 
$10.65\pm0.10$ & $10.73\pm0.07$ \\
\hline
\multicolumn{9}{c}{HgMn Stars}\\
\hline
$\beta$ Scl &221507 &12400 &3.90 &0.0: &27 & $9.60\pm0.10$ & 
$9.80\pm0.10$ & $9.70\pm0.07$ \\
36 Lyn & 79158 & 13700 & 3.65 & 2.0: & 49 & $9.90\pm0.10$ & 
$9.60\pm0.10$ & $9.75\pm0.07$ \\
$\upsilon$ Her &144206 &12000 &3.80 &0.6 & 11 & $10.20\pm0.05$ 
& $10.20\pm0.05$ & $10.20\pm0.04$ \\
HR 7361$^b$ &182308 &13650 &3.55 & 0.0 & 9 & $9.5\rightarrow9.8$ 
& $9.5\rightarrow9.8$ &  $9.65:$ \\
28 Her &149121 &11000 &3.80 &0.0 & 8 & $9.75\pm0.10$ & 
$9.90\pm0.10$ & $9.83\pm0.07$ \\
HR 7143 &175640 &12100 &4.00 &1.0 & 2 &  $10.12\pm0.05$ & 
$10.30\pm0.05$ & $10.21\pm0.04$ \\
46 Aql &186122 &13000 &3.65 &0.0 & 1 & $9.25\pm0.10$ & 
$9.45\pm0.10$ & $9.35\pm0.07$ \\
HR 7775 &193452 &10800 &3.95 &0.0 & 1 & $10.0\pm0.30$ & 
$9.50\pm20$ & $9.65\pm0.17$ \\
$\kappa$ Cnc &78316 &13500 &3.80 &0.0 & 6 & $9.85\pm0.15$ & 
$9.85\pm0.10$ & $9.85\pm0.08$ \\
53 Tau &27295 &12000 &4.25 &0.0 & 5 & $9.90\pm0.05$ & 
$10.20\pm0.10$ & $9.96\pm0.05$ \\
HR 7664$^b$ &190229 &13200 &3.60 &0.8 & 8 & $9.2\rightarrow9.7$ 
& $9.4\rightarrow9.8$ & 9.52: \\
$\phi$ Her &145389 &11650 &4.00 &0.4 &10 & $10.20\pm0.05$ & 
$10.38\pm0.05$ & $10.29\pm0.04$ \\
$\phi$ Phe &11753 &10700 &3.80 &0.5: & 13 & $9.80\pm0.15$ & 
$10.00\pm0.10$ & $9.94\pm0.05$ \\
$\nu$ Cnc &77350 &10400 &3.60 &0.1 &13 & $10.30\pm0.05$ & 
$10.10\pm0.10$ & $10.26\pm0.05$ \\
HR 2844 &58661 &13460 &3.80 &0.5: &30 & $10.00\pm0.10$ & 
$10.00\pm0.10$ & $10.00\pm0.07$ \\
33 Gem &49606 &14400 &3.85 &0.5: &22 & $9.50\pm0.10$ & 
$9.75\pm0.15$ & $9.58\pm0.08$ \\
$\mu$ Lep &33904 &12800 &3.85 &0.0 &18 & $9.65\pm0.10$ & 
$10.05\pm0.05$ & $9.97\pm0.05$ \\
HR 2676 &53929 &14050 &3.60 &1.0: &25 & $9.30\pm0.15$ & 
$9.70\pm0.30$ & $9.38\pm0.13$ \\
87 Psc &7374 &13150 &4.00 &1.5 &21 & $10.25\pm0.10$ & 
$10.30\pm0.10$ & $10.27\pm0.07$ \\
HR 6997 &172044 &14500 &3.90 &1.5 &34 & $9.72\pm0.05$ & 
$9.85\pm0.05$ & $9.79\pm0.04$ \\
HR 4072$^a$ &89822 &10650 &3.95 &1.0 &3.2 & $10.40\pm0.10$ & 
$10.40\pm0.10$ & $10.40\pm0.07$ \\
$\chi$ Lup$^a$ &141556 &10650 &4.00 &0.0 &2.0 & $10.76\pm0.04$ 
& $ 10.70\pm0.07$ & $10.74\pm0.02$ \\
$\iota$ CrB$^a$ &143807 &11000 &4.00 &0.2 &1.0 & $10.35\pm0.10$ 
& $10.42\pm0.05$ & $10.36\pm0.04$\\
112 Her$^a$ &174933 &13100 &4.10 &0.0: & 6 & $9.58\pm0.05$ &
$9.65\pm0.10$ & $9.60\pm0.04$\\
HR 1800$^a$ &35548 &11050 &3.80 &0.5 & 3 & $10.55\pm0.05$ & 
$10.45\pm0.05$ & $ 10.50\pm0.04$\\
\hline
\end{tabular}

$^a$Binaries with two spectra (parameters refer to the primary).  
Explicit allowances for dilution effects have been made in this 
work.\\
$^b$See text and Figs. 5 and 6 for discussion of HR7664 and 
HR7361.

\label{table:stars}
\end{table}

{\bf {Observations.}} The programme stars are listed in Table 
\ref{table:stars}.  Most of the observations were made with the Hamilton
\'{E}chelle Spectrograph at the Lick Observatory, with fwhm resolution
$R=46\,500$.  Two southern stars ($\xi$ Oct and $\beta$ Scl) were observed
with the UCLES on the Anglo-Australian Telescope with nearly identical
resolution.  Further details of the observations and reductions, and 
treatment of binary stars, are the same as in Jomaron et al (1999).

\vspace{0.3cm}

\begin{figure}[ht]
\begin{center}
\epsfig{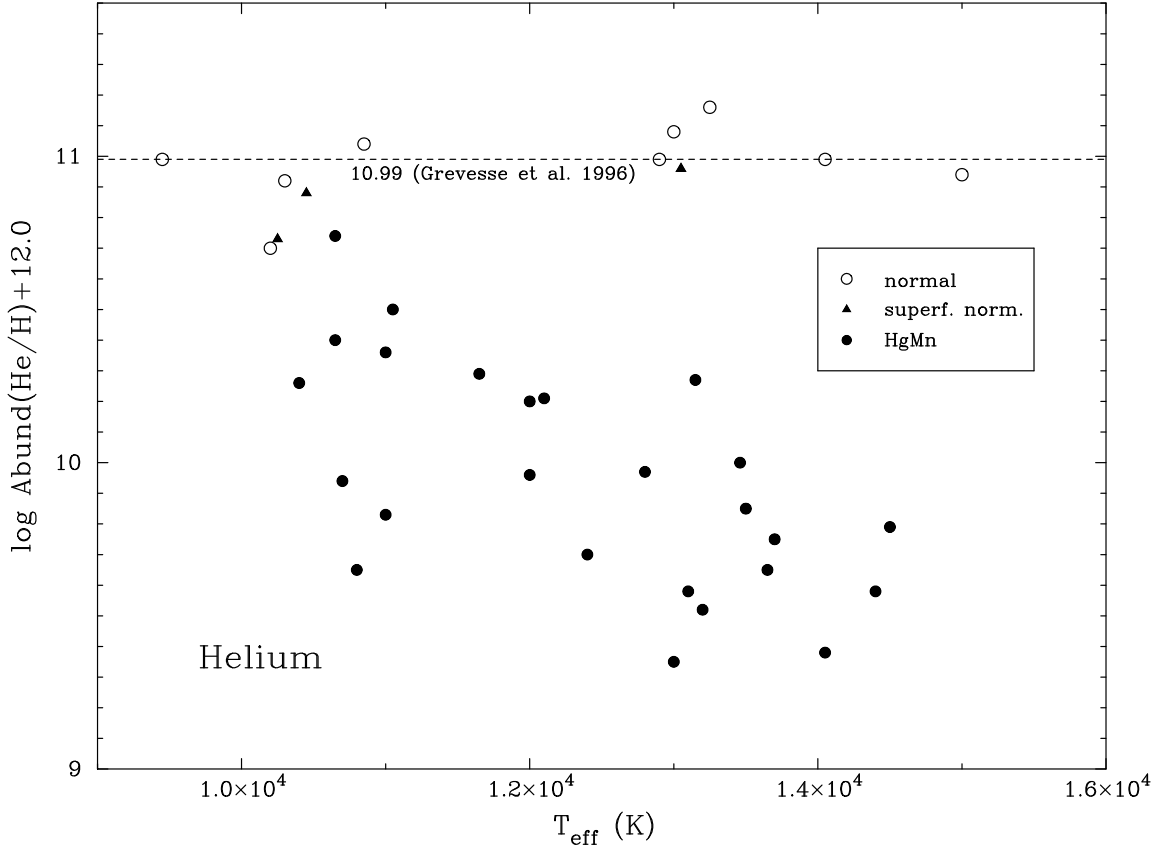}
\end{center}
\caption{Helium abundances in normal stars, superficially normal stars, 
and HgMn stars.  Points are averages of $\lambda$4026 and $\lambda$4471 
profile fits.  Typical errors (not shown) are $\pm0.06$\,dex.  There is 
more depletion of He in the hotter HgMn stars than in the cooler group.  
In two cases, where stratification of He is suspected, the values are the 
means of the best fits to line centres and wings (see text).}
\label{fig:abunds}
\end{figure}

{\bf {Abundance of He.}} Auer and Mihalas (1973) showed that the \hei\ 
lines
used in this work can be well-approximated by LTE models for B stars in
the temperature range below 15000\,K. The spectrum synthesis code {\sc
uclsyn} was used to calculate the abundance of He for 9 normal stars, 3
superficially-normal stars, and 25 HgMn stars.  The results are shown in
Table~\ref{table:stars} and Fig~\ref{fig:abunds}. It is clear from this
figure that all HgMn stars in this sample are deficient in He by 0.3 dex
or more (usually much more).  It also appears that the hotter HgMn stars
have stronger He deficits than the cooler HgMn stars.  Dividing them into
12 cooler and 13 hotter HgMn stars yields $\log N(\mathrm {He}) = 10.20
\pm 0.09$ for the cooler group and $\log N(\mathrm {He}) = 9.72 \pm 0.07$
for the hotter group.  The statistical difference is highly significant.

\vspace{0.5cm}

\begin{figure}[ht]
\begin{center}
\epsfig{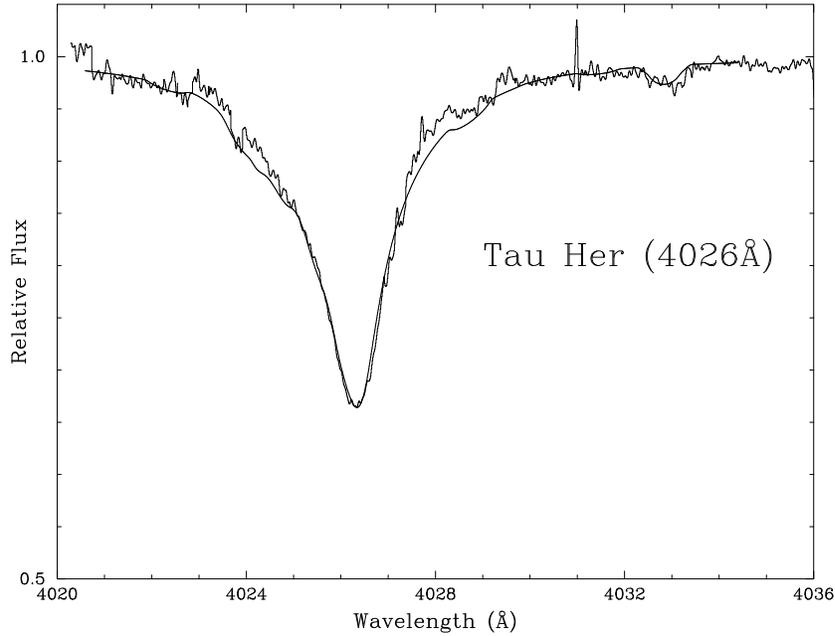}
\end{center}
\caption{The line profile fit for $\tau$ Her, a
normal B5\,IV star with \teff\ = 15,000\,K, $\log g = 3.95$, and
derived He abundance 10.94.}
\label{fig:tauher}
\end{figure}

\begin{figure}
\begin{center}
\epsfig{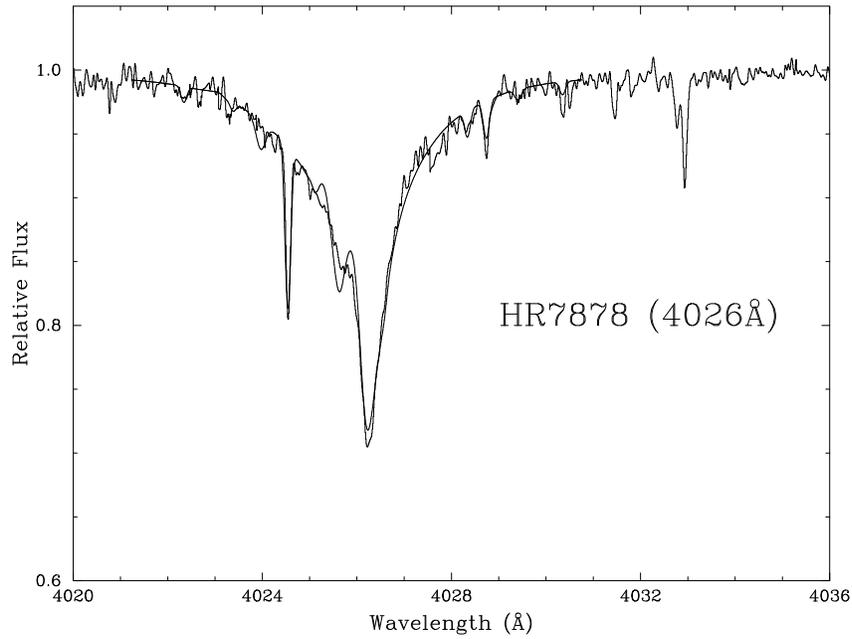}
\end{center}
\caption{The line profile fit for HR7878
(B8\,IIIp), a superficially normal star with \teff\ = 13,050,
$\log g = 3.85$, with He abundance 10.85 for $\lambda$4026.}
\label{fig:hr7878}
\end{figure}

\begin{figure}
\begin{center}
\epsfig{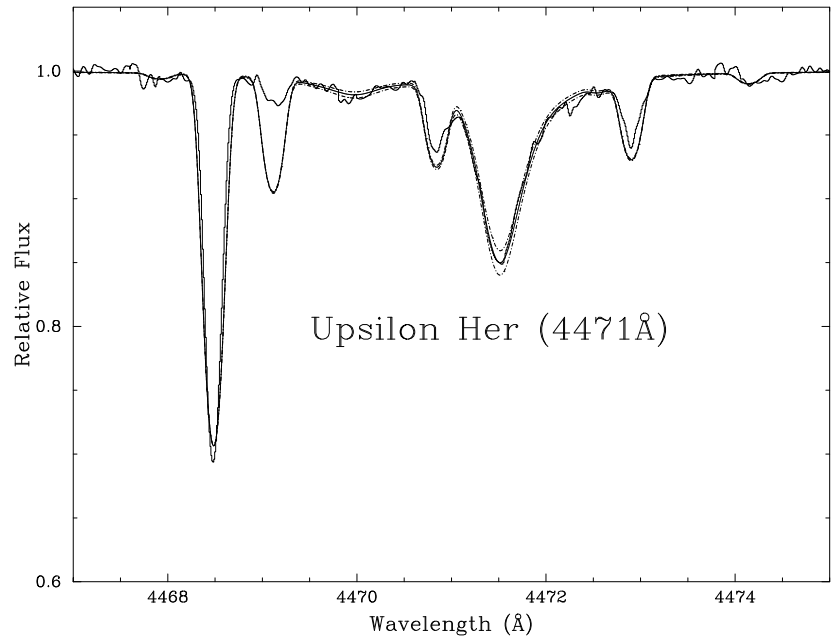}
\end{center}
\caption{Best fit for HgMn star $\upsilon$ Her
$\lambda$4471, \teff\ = 12,000\,K, $\log g = 3.80$, with He
abundance 10.20.  For illustration, the dash-dot lines indicate
the profiles for changes of abundance of $\pm0.08$ dex, larger
than the error estimate in this work.}
\label{fig:upsher}
\end{figure}

Examples of fits to normal and HgMn stars are shown in 
Figs~\ref{fig:tauher},~\ref{fig:hr7878}~\&~\ref{fig:upsher}.

\begin{figure}
\begin{center}
\epsfig{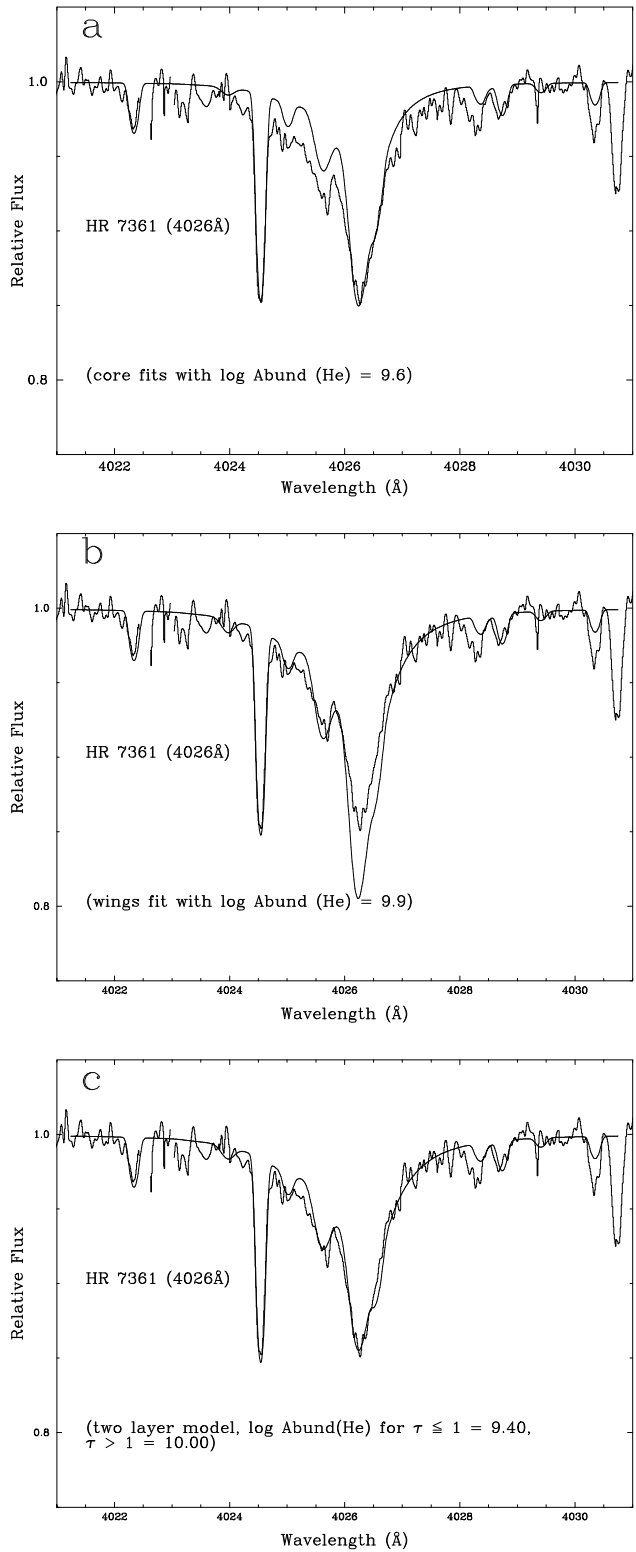}
\end{center}
\caption{The centre or core (a) of $\lambda$4026 \hei\ fits a very low He 
abundance in the HgMn star HR7361.  The wings can only be fit with a 
higher abundance (b), but then the core is a bad fit.  A model with a 
higher abundance of He below $\tau=1$ (c) fits both core and wings 
reasonably well.  This indicates that He is stratified, with more 
depletion in the upper photosphere than in the deeper photosphere.}
\label{fig:hr7361}
\end{figure}

\begin{figure}
\begin{center}
\epsfig{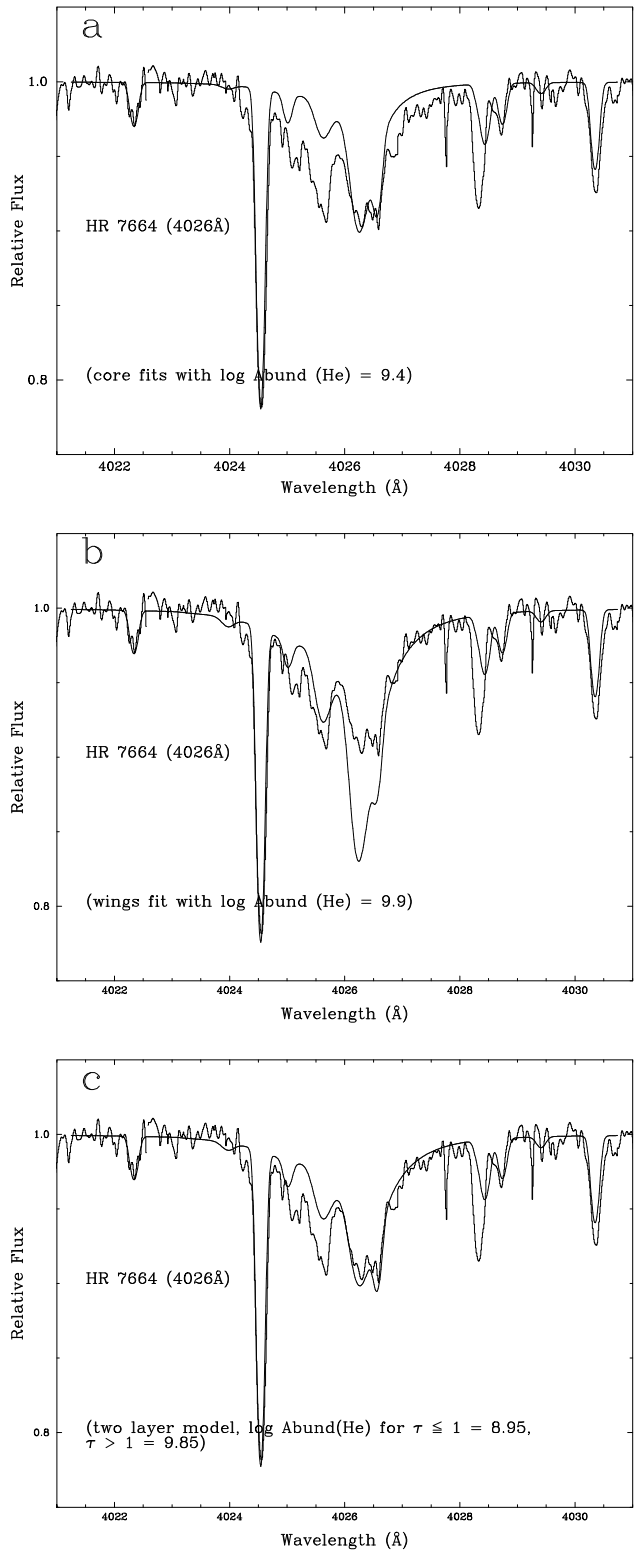}
\end{center}
\caption{The HgMn star HR7664 has similar fitting problems to HR7361 
shown in Fig.~\ref{fig:hr7361}.  The core (a) of $\lambda$4026 
\hei\ fits a low He log abundance, 9.4.  The wings require a higher value 
9.9 (b), but then the core is a bad fit.  A model with 
a higher abundance of He below $\tau=1$ (c) fits both core and wings
reasonably well.  Again, this shows that He is stratified, with more
depletion in the upper photosphere than in deeper layers.}
\label{fig:hr7664}
\end{figure}

Typical errors of single determinations by fitting one line were 
$\pm0.05 - 0.10$ dex, and consistency between $\lambda$4026 and 
$\lambda$4471 was of similar quality.  Blends which occurred 
within the profile were modelled.

\vspace{0.3cm}

{\bf {Depth Dependent He Abundances.}} In nearly all cases, the
profile fits to wings and centres of the two lines were fully
consistent, indicating that the modelling assumption of uniform
fractional abundance of He with depth was a good one.  However,
two HgMn stars, HR7361 (Fig.~\ref{fig:hr7361}) and HR7664 
(Fig.~\ref{fig:hr7664})  could 
not be
fit to one abundance.  In both cases the centre gave a good fit
only for abundances about 0.3-0.5 dex lower than the fit in the
wings.  A model of the He abundance in these two stars with an
enhanced abundance below $\log\tau = 1.0$ produced much more
satisfactory fits.  Although the solutions in Figs.~\ref{fig:hr7361}c and 
\ref{fig:hr7664}c are
not unique, as the actual depth distribution is probably more
complicated, they are indicative of the fact that He must be
considerably more depleted in the higher photosphere than in
deeper layers, although it is also depleted there as well.  It
appears that He has left a clear trace of its downwards diffusion
through the \heii\ convection zone in these two cases.

\vspace{0.3cm}

{\bf {Acknowledgements.}}  I am grateful to Prof. J. S. Miller,
Director of Lick Observatory for observing time for this work,
and to Mr. D. Stansall, MSci, for his excellent efforts on He
during a supervised undergraduate research project at University
College London.

\vspace{0.3cm}

\noindent
{\bf{References:}}

\noindent 
Auer L.H., Mihalas, D., 1973, ApJS 25, 433\\ 
Barnard, A.J., Cooper, J., Shamey, L.J., A\&A, 1, 28\\
Barnard, A.J., Cooper, J., Smith, E.W., 1974, JQSRT, 14, 1025\\ 
Barnard, A.J., Cooper, J., Smith, E.W., 1975, JQSRT, 15, 429\\ 
Grevesse, N., Noels, A., Sauval, A.N., 1996, ASP Conf Series, 99, 
117\\
Jomaron, C.M., Dworetsky, M.M., Allen, C.S., 1999, MNRAS, 303, 
555\\ 
Michaud, G., 1986, in Cowley, C.R., Dworetsky, M.M.,
M\'{e}gessier, C., eds, Upper Main Sequence Stars with Anomalous
Abundances, IAU Coll. 90, D. Reidel, Dordrecht, 459\\ 
Michaud, G., Martel, A., Montmerle, T., Cox, A.N., Magee, N.H., 
Hodson, S.W., 1979, ApJ, 234, 206\\ 
Shamey, L.J., 1969, PhD Thesis, U. Colorado\\ 
Smith, K.C., Dworetsky, M.M., 1993, A\&A, 274,335\\ 

\end{document}